\documentclass[article,structabstract]{aa}
\usepackage{graphicx}

\usepackage{amsmath}
\usepackage{txfonts}

\usepackage{multirow}
\usepackage{xspace}
\usepackage{verbatim} % for comments
\usepackage{natbib}
\bibpunct{(}{)}{;}{a}{}{,} % to follow the A&A style
\newcommand{\um}{\ensuremath{\rm \mu m}\xspace}

\newcommand{\geo}{\ensuremath{\omega_{\rm disk}}\xspace}
\newcommand{\afilt}{\ensuremath{\omega_{\rm filt}}\xspace}
\begin{document}
\title{Why circumstellar disks are so faint in scattered light: The case of HD 100546}
\titlerunning{Scattering paper}
 \author{Gijs D.~Mulders\inst{1,2}
\and Michiel~Min\inst{1}
\and Carsten~Dominik\inst{1,3}
\and John H.~Debes \inst{4}
\and Glenn Schneider\inst{5}
}

\institute{
%1
Astronomical Institute ``Anton Pannekoek'', University of Amsterdam,
 PO Box 94249, 1090 GE Amsterdam, The Netherlands
\and %2
SRON Netherlands Institute for Space Research, PO Box 800, 9700 AV,
Groningen, The Netherlands
\and %3
Department of Astrophysics/IMAPP, Radboud University Nijmegen,
P.O. Box 9010 6500 GL Nijmegen The Netherlands
\and %4
Space Telescope Science Institute, Baltimore, MD 21218, USA
\and %5
Steward Observatory, The University of Arizona, 933 North Cherry Avenue, Tucson, AZ 85721, USA
} % end institutes

%\date{ }
\offprints{G.D.Mulders, \email{\bf{mulders@uva.nl}}}

\abstract{%context
  Scattered light images of circumstellar disks play an important role in characterizing the planet forming environments around young stars. The characteristic size of the scattering dust grains can be estimated from the observed brightness asymmetry between the front and back side of the disk, for example using standard Mie theory. However such models often overpredict their brightness by one or two orders of magnitude, and have difficulty explaining very red disk colors.
 }{%aims
   We aim to develop a dust model that explains simultaneously the observed disk surface brightness, colors and asymmetry in scattered light, focussing on constraining grain sizes.
}{%methods
  We use the 2D radiative transfer code MCMax with anisotropic scattering to explore the effects of grain size on synthetic scattered light images of circumstellar disks. We compare the results with surface brightness profiles of the protoplanetary disk \object{HD 100546} in scattered light at wavelengths from 0.4 to 2.2 micron.
}{%results
We find that extreme forward scattering by micron sized particles lowers the \textit{effective} dust albedo and creates a faint and red disk that \textit{appears} only slightly forward scattering. For the outer ($\gtrsim$100 AU) disk of HD 100546 we derive a minimum grain size of 2.5 micron, likely present in the form of aggregates. Intermediate sized grains are too bright, whereas smaller grains are faint and scatter more isotropically, but also produce disk colors that are too blue.
}{%conlusions
Observed surface brightness asymmetries alone are not sufficient to constrain the grain size in circumstellar disks. Additional information, such as the brightness and colors of the disk are needed to provide additional constraints.
}

\maketitle

\begin{figure*}
  \includegraphics[width=\textwidth]{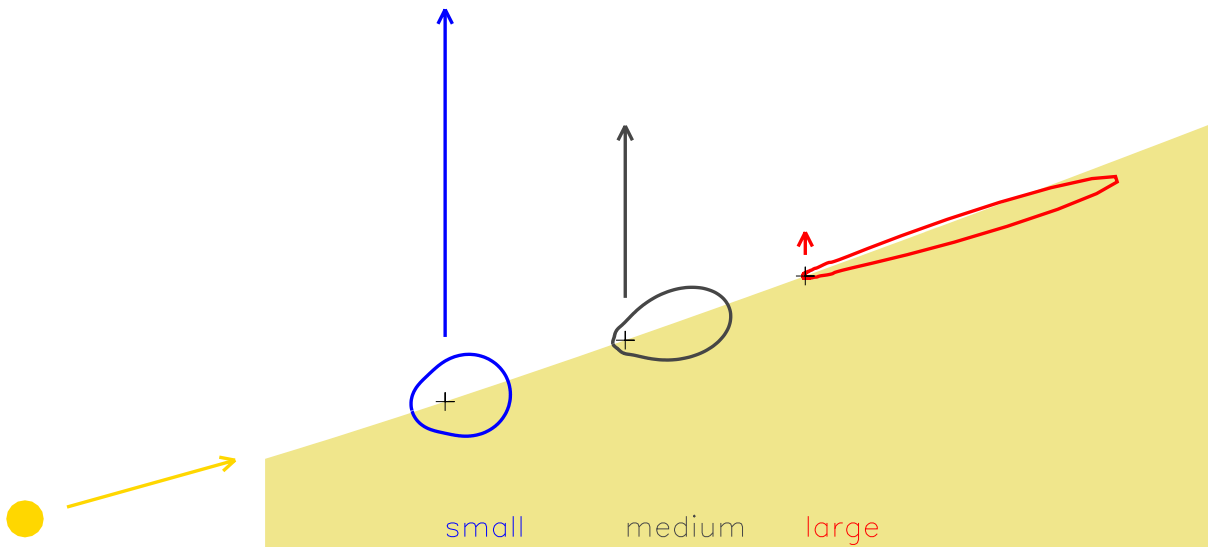}
  \caption[]{Visualization of anisotropic scattering properties of different dust species. Displayed are: a small particle in the Rayleigh limit ($2\pi a < \lambda $) in blue, an intermediate sized particle ($2\pi a \sim \lambda $) in gray and a large particle ($2\pi a > \lambda $) in red. The angle between the disk surface and incident stellar light is 3 degrees, equal to the flaring angle minus aspect ratio of the surface at 200 AU. The phase functions at 0.6 \um are plotted in polar coordinates, and normalized such that surfaces are equal (square root of intensity). The arrows indicate the direction and magnitude of light scattered towards the observer.
    \label{fig:cartoon}}
\end{figure*}

\section{Introduction}
Protoplanetary disks are thought to be the main sites of planet formation, and spatially resolved scattered light images play an important role in revealing what happens before and during the birth of planetary systems. With superior diffraction-limited resolution compared to thermal emission at longer wavelengths, scattered light images in general - and in particular those from the Hubble Space Telescope - have provided us with some of the most detailed images of protoplanetary disks so far \cite[e.g.][]{2001AJ....122.3396G}. Recent advances in adaptive optics and data reduction techniques have greatly improved the efficacy of ground-based data \citep[e.g.][]{2010ApJ...718L..87T,2011ApJ...729L..17H, 2011ApJ...738...23Q}, while future observatories such as ALMA will achieve similar spatial resolution at (sub)millimeter wavelengths.

Scattered light images offer important information on the overall geometry of protoplanetary disks and contain possible signposts of planets. Large scale spiral structures have been observed, some triggered by a close encounter from outside \citep{2003AJ....126..385C,2005AJ....129.2481Q}, others more likely by a (planetary) perturber from within \citep[e.g.][]{2001AJ....122.3396G}. Annular gaps and inner holes are also observed, and are likely carved by orbiting protoplanets \citep{2007ApJ...665.1391G,2010ApJ...718L..87T,2011ApJ...729L..17H}.

Apart from the macroscopic signposts at the end of the (giant) planet formation process, scattered light imaging can also shed light on the earliest stages of planet formation: the growth of dust grains. The main diagnostics for this are:
\begin{enumerate}
  \item Brightness asymmetries between near and far side of the disk. In an inclined disk, the near and far side of the disk surface are seen under a different angle with respect to the star. If the dust scatteres anisotropically, a brightness asymmetry is observed. Larger dust grains are more forward scattering, and will show a stronger brightness asymmetry. The observed values for the asymmetry parameter g in different disks are between $\sim$0.15 \citep{2007ApJ...665..512A} and $\sim$0.8 \citep{2008A&A...489..633P}, ranging from nearly isotropic to slightly forward scattering.
  \item Disk surface brightness or dust albedo. The fraction of scattered light at a specific wavelength. In general, disks are observed to be much fainter that predicted on basis of their grain size. A sample of Herbig stars imaged by \cite{2010PASJ...62..347F} shows fractional luminosities in scattered light on the order of a few percent or less. This is much too low for non-Rayleigh scattering particles, which typically have an albedo of $\sim$ 0.5. In debris disks, which are optically thin, albedos on the order of 0.05...0.1 are derived from observations \citep{2005Natur.435.1067K,2010AJ....140.1051K,2011AJ....142...30G}, while an albedo of 0.5 is predicted on basis of the grain radius of $\sim$1 \um derived from asymmetries in the disk images.
  \item Disk color, caused by the wavelength dependence of the albedo. Dust grains in the interstellar medium have a grey scattering color at optical to near-infrared wavelengths \citep[e.g.][]{1995RMxAC...1..201W}, whereas more evolved objects such as debris disks \citep[e.g.][]{2006AJ....131.3109G,2008ApJ...673L.191D}, comets \citep{1986ApJ...310..937J} and Kuiper Belt objects \citep[e.g.][]{1996AJ....112.2310L} can have redder colors, indicating that larger than ISM grains are present \citep[e.g.][]{2012A&ARv..20...52W}. Most protoplanetary disks have grey colors \citep[e.g.][]{2010PASJ...62..347F}, though some disks exhibit redder colors: \object{HD 141569} \citep{2003AJ....126..385C}, \object{GG Tau} \citep{2005AJ....130.2778K} and \object{HD 100546} \citep{2007ApJ...665..512A}.
\end{enumerate}

To extract the grain size from scattered light images, these three aspects (moderate brightness asymmetries, low albedos, neutral to red colors) have to be modelled simultaneously. This is challenging, because small dust grains have low albedos but very blue colors. Large grains have neutral to red colors but also high albedos. We call this ``the color and brightness problem of scattered light images''. 

To solve this problem, we have studied the effects of anisotropic scattering on the observed brightness and colors. Particles that are large enough become extremely forward scattering. The bulk of the scattered light is concentrated in the forward peak - which is outside the range of observed angles for a disk that is not observed edge-on (Fig. \ref{fig:cartoon}). The resulting few percent are scattered towards the observer, resulting in a low 'effective' albedo \citep{2003A&A...408..161D}. Because the observer of an inclined disk sees only a part of the phase function outside of the forward scattering peak, the disk image does not appear to have a strong brightness asymmetry, and can even \textit{appear} backward scattering \citep{2010A&A...509L...6M}. We will describe this effect in more detail in section \ref{sec:method}.

We have applied this strong forward-scattering dust model to the well-studied protoplanetary disk of HD 100546, for which scattered light images are available from 0.4 to 1.6 $\mu$m \cite{2001A&A...365...78A,2007ApJ...665..512A}, and we present a new image at 2.22 $\mu$m in section \ref{sec:obs}. After constructing a geometric model of the disk in section \ref{sec:diskmodel}, we derived the grain size from the observations in sections \ref{sec:hg} and \ref{sec:color}. We will also discuss how the derived grain size compares to those of other size indicators in the discussion.

\section{Scattering by protoplanetary dust}\label{sec:method}

In this section we will briefly discuss the basis of the scattering properties of protoplanetary dust \citep{Mie:1908tn,VanDerHulst,1983asls.book.....B} and its influence on the observed brightness and color of scattered light images. 

\begin{figure*}
  \includegraphics[width=\textwidth]{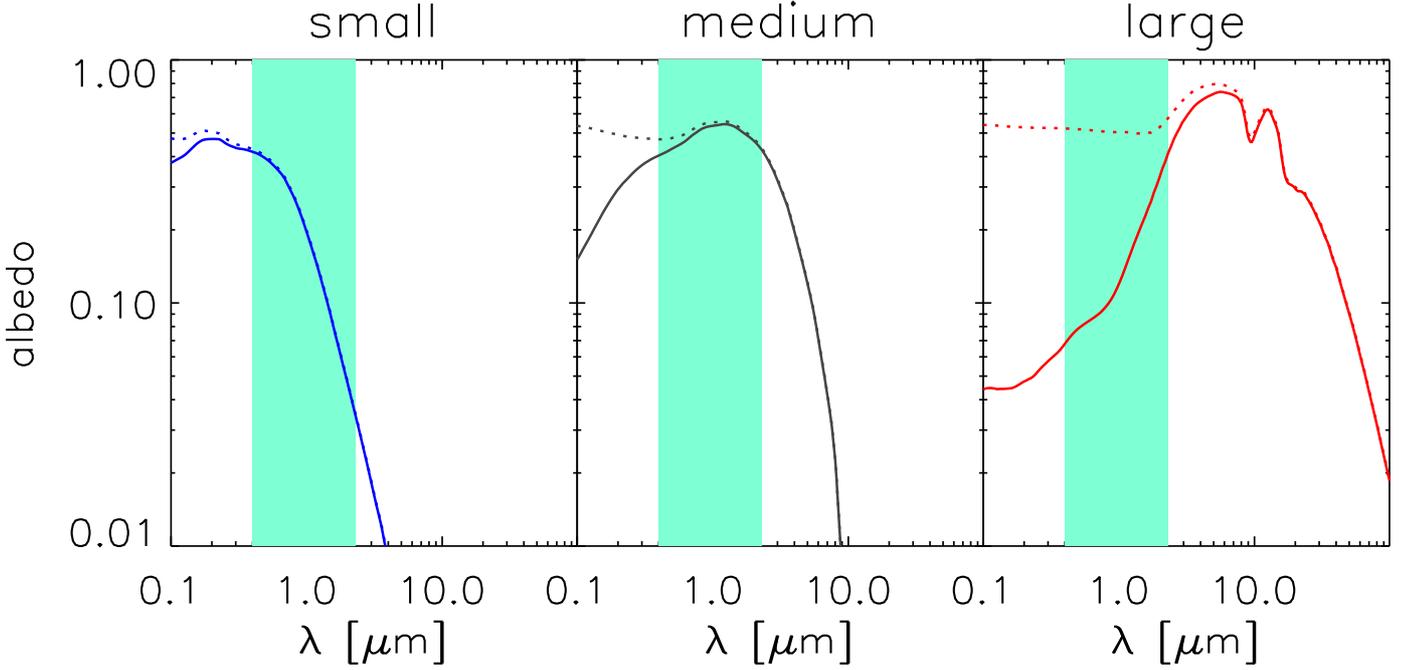}
  \caption[]{Real versus effective albedo of particles of different sizes. The effective albedo (solid line) is defined as the fraction of light scattered outside of the forward ten degrees times the real albedo, see text (dotted line). The green area denotes the range of observed wavelengths by the Hubble scattered light images. The small, medium and large panels correspond to a particle size of 0.08 \um, 0.25 \um and 2.5 \um.
    \label{fig:effective}}
\end{figure*}

\subsection{Single scattering albedo}
The fraction of light that is scattered by a dust grain in all directions is given by the single scattering albedo, $\omega$. For particles much smaller than the wavelength the albedo is very low since most light is absorbed by the particle. When the size of the particle increases, the single scattering albedo also increases. For particles much larger than the wavelength of incident radiation, the total extinction cross section of the particle is comprised of three components: absorption ($C_\mathrm{abs}$), reflection/refraction ($C_\mathrm{ref}$), and diffraction ($C_\mathrm{diff}$).

The absorption and reflection/refraction cross sections come directly from geometrical optics and sum up to the geometrical shadow of the particle. The diffraction cross section comes from the distortion of the wavefront caused by the particle and is also equal to the geometrical shadow of the particle. It is than directly derived that the single scattering albedo of large particles is, 
\begin{equation}\label{eq:albedo}
\omega=\frac{C_\mathrm{scat}}{C_\mathrm{ext}}=\frac{C_\mathrm{ref}+C_\mathrm{diff}}{C_\mathrm{abs}+C_\mathrm{ref}+C_\mathrm{diff}} \gtrsim 0.5.
\end{equation}

\begin{figure}
  \includegraphics[width=\linewidth]{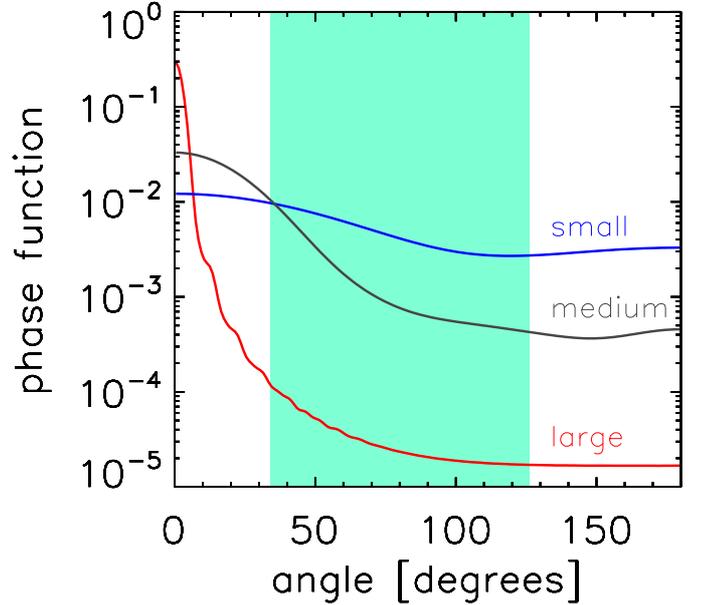}
  \caption[]{Phase function of a large dust grain (2.5 \um, red line), an intermediate size dust grain (0.25 \um, grey line) and a small dust grain (0.08 \um, blue line) at 0.6 \um. The green area marks the range of observed angles for HD 100546, between 34\degr{} and 126\degr. $\theta=0^o$ refers to forward scattering and $\theta=180^o$ to backward scattering.
    \label{fig:phase}}
\end{figure}

\begin{figure*}
  \includegraphics[width=\textwidth]{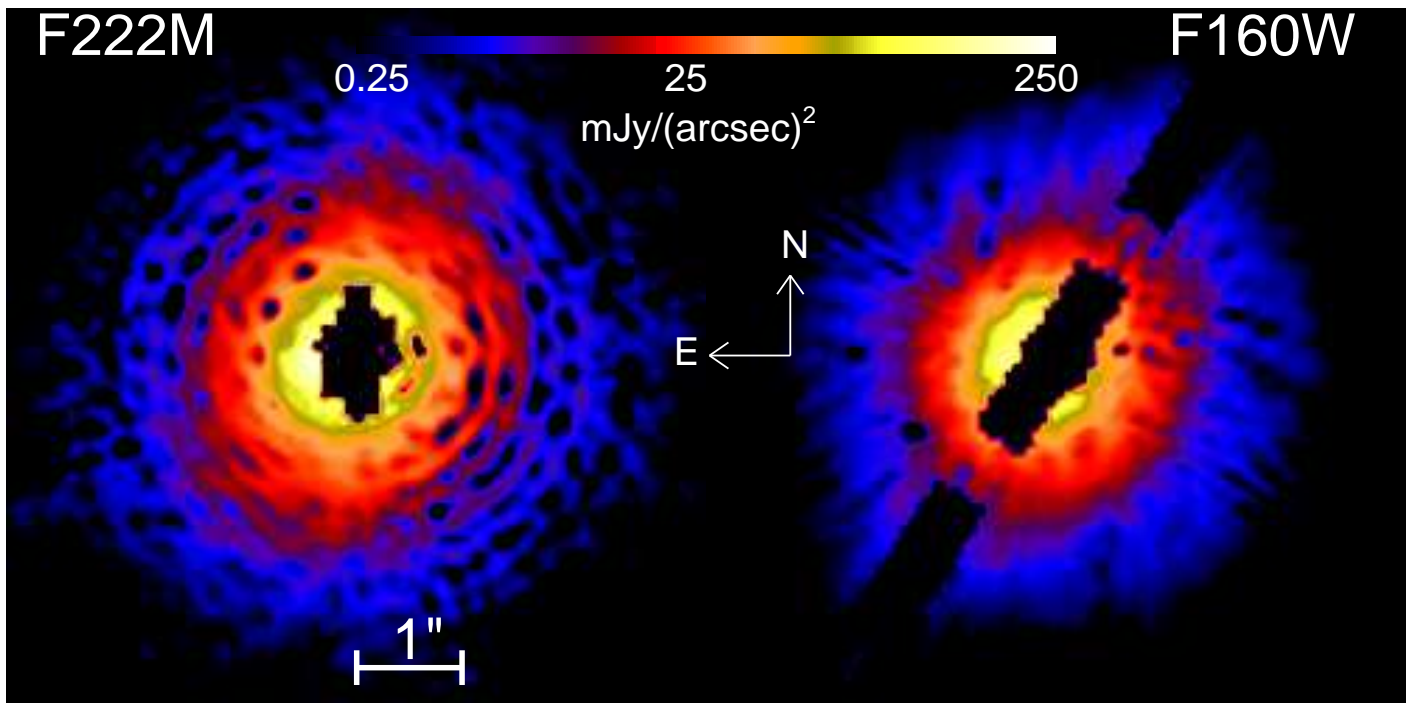}
  \caption[]{PSF subtracted NICMOS images at 1.6 and 2.22 \um. The black region in the F160W image is masked out due to a strong diffraction spike.
    \label{fig:image}}
\end{figure*}

\subsection{The effective albedo and the color of scattered light}
However, the single scattering albedo is an angle-integrated quantity, and care should be taken when scattering becomes increasingly more anisotropic. In the limit of Rayleigh scattering -- where particles are much smaller than the observed wavelength -- the phase function ($F_{11}(\theta)$) of scattering is close to isotropic (Fig \ref{fig:phase}). For particles larger than the wavelength the scattering becomes anisotropic and heavily peaked towards the forward direction when the particle size increases.

Light which is scattered in the forward direction by a particle in the surface layer of the disk is scattered into the disk, and will not be detected in scattered light images (see Fig.~\ref{fig:cartoon}, large particle). Therefore, it is clear that for particles that scatter predominantly in the forward direction the \textit{effective} scattering cross section, i.e. the amount of light that is actually scattered into our line of sight, is much smaller than would be expected from the \textit{angle integrated} single scattering albedo.
Therefore we define an effective albedo which only takes into account the part of the phase function that can actually be observed between angles $\theta_1$ and $\theta_2$:
\begin{equation}
\omega_\mathrm{eff}=\frac{2\,\omega \int_{\theta_1}^{\theta_2} F_{11}(\theta)\sin\theta d\theta}{(\cos\theta_1-\cos\theta_2)\int_{0}^{\pi} F_{11}(\theta)\sin\theta d\theta}.
\end{equation}
For large particles this effective albedo can be smaller than 0.5 when the phase function is sufficiently peaked towards the forward scattering direction. 

In Fig.~\ref{fig:effective} we plot for three different grain sizes the effective albedo for arbitrarely\footnote{Cutting out the forward 10 degrees is sufficient to illustrate the main effect of not observing the forward scattering peak} $\theta_1=10^o, \theta_2=180^o$ together with the total albedo. For the small grains the scattering is quite isotropic (see Fig. \ref{fig:phase}) and the effective and total albedo are almost the same. The color of scattered light for these grains is always grey to slightly blue.

The phase function of larger particles is more anisotropic, making the effective albedo quite low, even below the limit for the \textit{single scattering albedo} (Eq. \ref{eq:albedo}). When going to longer wavelengths the phase function becomes more isotropic and the effective albedo increases. This increase of effective albedo with wavelength will be reflected in the observed color of the scattered light images, producing reddish colors.

\section{Application to HD 100546}\label{sec:application}
To test if exteme forward scattering can explain the observed faintness of protoplanetary disks relative to their host stars \citep{2010PASJ...62..347F}, as well as their color indices and observed brightness asymmetries, we have compared our dust model to scattered light images of a well-studied Herbig star, \object{HD 100546}. Its disk has been imaged by the Hubble Space Telescope in scattered light over a broad wavelength range of 0.4 \um to 1.6 \um \citep{2001A&A...365...78A,2007ApJ...665..512A}. We will extend this wavelength range using a new NICMOS image at 2.2 $\um$. This large wavelength range allows us to compare both brightness and colors. We will describe these observations in the next section.

\subsection{Observations}\label{sec:obs}
We took HST/NICMOS coronagraphic images (0.3" radius image plane obscuration, camera 2 image scale 75.8 mas/pixel) of HD~100546 and the PSF reference star, HD~109200 with the  F222M filters (central $\lambda=2.22~\um$, $\lambda/\Delta\lambda\sim10$) on 16 March 2005 as part of Program GO 10167 (PI:Weinberger).  The observations include long exposures with the stars underneath the coronagraphic spot for high contrast imaging at two different spacecraft orientations and direct images of both stars outside the coronagraphic hole with short exposures for point source photometry.   The instrumentally calibrated and reduced images discussed in this paper were created from the raw NICMOS {\it multiaccum} exposures following the processing methodolgy  described by \S3 of \cite{2005ApJ...629L.117S} and references therein.

For photometric analysis, each calibrated direct image was used separately to independently determine the total photometry of the star and empirically determine the uncertainties in each filter band.  The three images for each star and in each filter were located at different positions on the detector.

We used a median combination of the three dither points to create a final image of each star to derive a ratio for scaling and for photometry of HD~100546.  We used a 20 pixel radius circular aperture to determine the photometry.  The background in the images is zero, so no background annulus was used.  The individual dither points were used to get a rough estimate of the uncertainty in the ratios and photometry.  At F222M we measure a total $F_\nu$=5.4$\pm$0.1$~Jy$, and a scaling with the PSF reference of 0.89$\pm$0.04.

In order to determine the best subtraction we minimized a chi-squared metric on a region of the target image dominated by the star's diffraction spikes.  We assumed that good subtraction of the diffraction spikes in a region uncontaminated by the disk corresponded to the best subtraction of the PSF within the region of interest \citep{2001AJ....121..525S}.  We iteratively created subtractions for combinations of scaling and pixel offsets until we found an image that produced the lowest chi-squared measure.  We searched within $\pm$1~pixel to find the best x and y pixel offsets. 

To quantify the systematic effects on the photometry, we repeated the subtractions varying the PSF scalings and offsets by $\pm$1 $\sigma$ from the minimum chi-square solution found above. Using an elliptical photometric aperture matched to the estimated inclination of the disk and that extended from between 0\farcs5 to 4\arcsec, we found the standard deviation in the disk flux densities from this suite of subtractions.  In F160W and F222M, the total measured flux of the disk at both spacecraft orientations matched to within the uncertainties.  We then propagated this systematic uncertainty into the total uncertainty in the flux density of the disk per pixel.  Subtracted images were then geometrically corrected for the slight optical distortion of the NICMOS camera 2 at the coronagraphic focus.  We used the x-direction pixel scale of 0\farcs07595/pixel and the y-direction pixel scale of 0\farcs07542/pixel to create an image with pixels that have the y-direction plate scale in both directions.  The geometrically corrected images were rotated about the position of the occulted star to a common celestial orientation using the rotation centers given by the flight software in the raw data file headers. The final result is show in figure \ref{fig:image}.

Additional observations of HD~100546 with NICMOS were performed as a part of the GO program 9295 \citep{2001A&A...365...78A}.  Observations with NICMOS in the F160W filter were recovered from the archive, reduced in the same manner as the medium bandwidth image, and median combined.  Archival PSF reference stars for the images were subtracted from the target observations.  The F160W reference was HD~106797 as used in \citet{2001A&A...365...78A} but we followed the above procedure for subtraction as with the F222M filter.  Figure \ref{fig:image} shows the resulting PSF subtracted NICMOS images of the HD~100546 disk used in this work.

HD~100546 was also observed with the HRC coronagraph on ACS in the F435W and F814W filters \citep{2007ApJ...665..512A}.  The final reduced images were kindly provided to us by D. Ardila.  For the ACS observations, no independent estimate on the uncertainties in flux scaling for the disks was performed.  We assumed that for both instruments this affect was $\sim$5\%, comparable to what we calculated for our medium bandwidth filter observations.

\subsection{Surface brightness profile and disk albedo}\label{sec:profile}
\begin{figure*}
  \includegraphics[width=\linewidth]{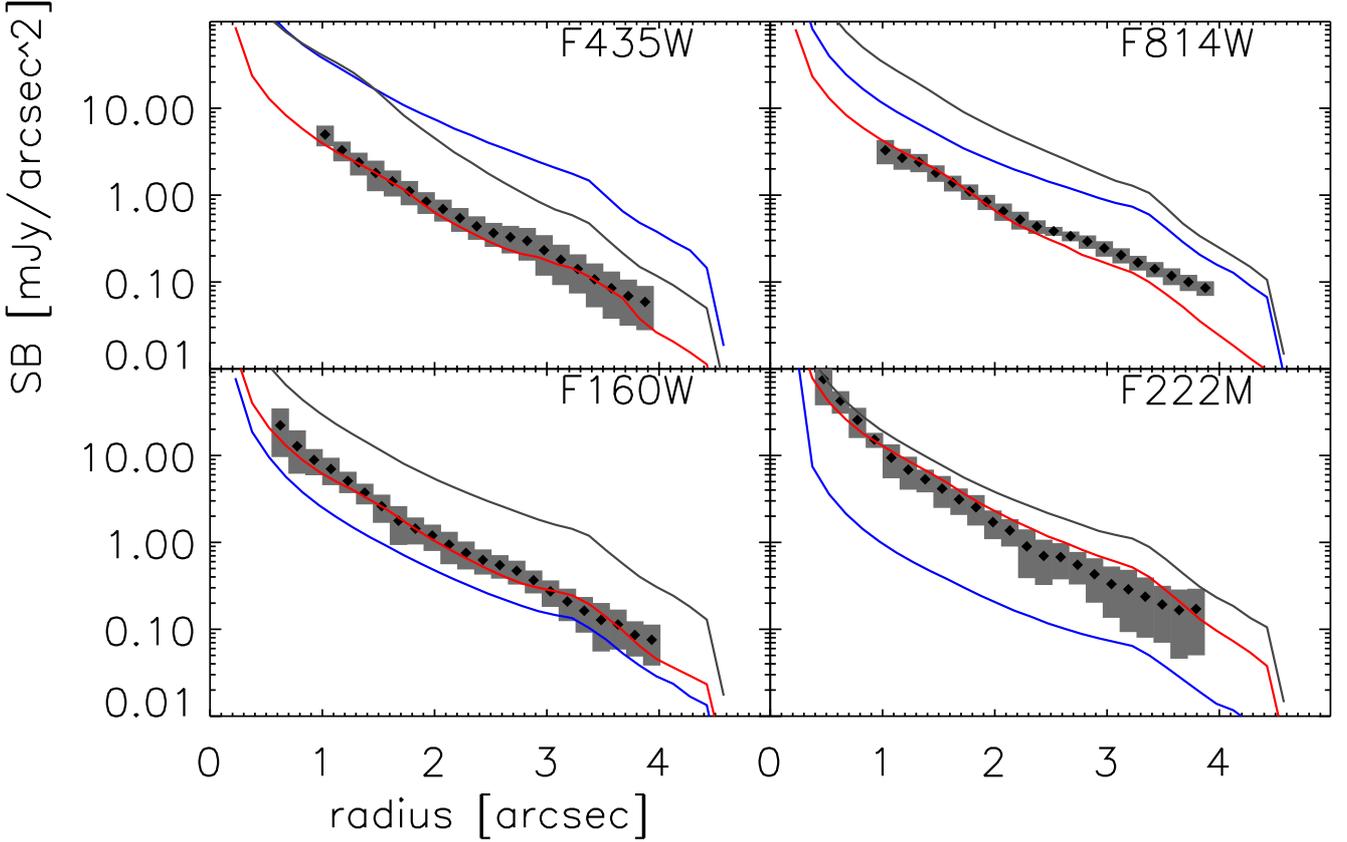}
  \caption[]{Radial surface brightness profiles of scattered light images of HD 100546 (diamonds, grey area indicates systematic and measurement errors) and disk models with: A large dust grain (2.5 \um, red line), an intermediate size dust grain (0.25 \um, gray line) and a small dust grain (0.075 \um, blue line).
    \label{fig:sbp}}
\end{figure*}

The surface brightness profiles at 0.4, 0.8, 1.6 and 2,2 \um are shown in figure \ref{fig:sbp}. They were constructed by taking the median surface brightness over an elliptical annulus whose shape corresponds\footnote{Although the flux coming from a certain distance from the star deviates from an ellipse if the disk is flared, we don't expect this to influence our analysis as long as we compare observed and model images in the same way.} to a disk at a position angle of 145\degr{} and inclined at 46\degr{} - the average inclination inferred by \cite{2000A&A...361L...9P} and \cite{2007ApJ...665..512A}. To be able to compare the surface brightness profiles of images taken at different spatial resolution, the width of the annuli is taken to be 0.15\arcsec, the diffraction limit of the longest wavelength filter. For the uncertainties we used local estimates at the major and minor axes of the standard deviation of counts within smaller apertures.  Missing data were treated as NAN values and were not included in the median.

To measure the reflectivity of the disk observed in each filter, we define a geometric albedo \geo \textit{for the disk}, which is the fraction of (star)light scattered towards the observer at a specific radius $r$. It is defined as:
\begin{equation}\label{eq:geo}
\geo (r) = \frac{{\rm SB}(r)}{(F_*+F_{\rm inner}) / 4 \pi r^2 }
\end{equation}
where SB$(r)$ is the surface brightness of the disk in scattered light, $F_*$ is the flux density from the star and $F_{\rm inner}$ is the flux density from the inner ($\lesssim$1 AU) disk, both measured from the unresolved SED\footnote{Note that at near-infrared wavelengths, the thermal emission from the inner disk is much brighter than that of the star. The light that we observe in the longest filters is therefore dominated by scattered inner disk light, rather than scattered starlight.}. Note that this 'disk albedo' \geo depends on both the dust albedo \textit{and} the geometry of the disk (i.e., its flaring angle).

\begin{figure}
  \includegraphics[width=\linewidth]{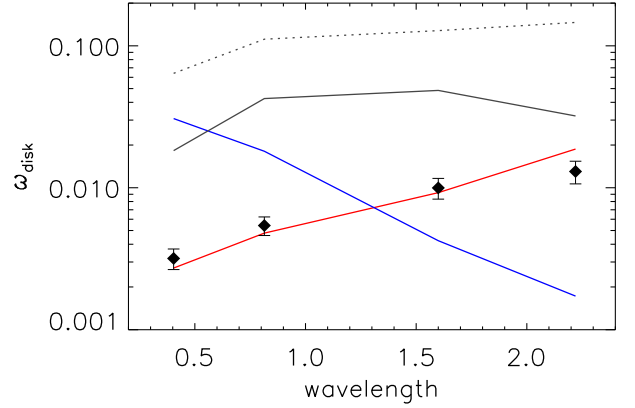}
  \caption[]{Geometric albedo \geo of HD 100546 in the different filters. The geometric albedo is a measure of the fraction of light scattered by the disk (see eq. [\ref{eq:geo}] and text for definition). Displayed are both observations (diamonds with systematic and measurement errors) and the same models as figure \ref{fig:sbp}: A large, intermediate and small grain (red, gray and blue solid line respectively). A model with intermediate sized grains with an albedo close to one is also displayed (gray dotted line).
    \label{fig:disk_color}}
\end{figure}

To be able to quickly compare disk albedos at different wavelengths for both the data and models, we define a geometric albedo per filter \afilt at a typical radius of $2\arcsec$. We obtain \afilt by fitting a powerlaw of the form $\afilt \cdot (r)^{-q}$ to \geo$(r)$ between 1.0\arcsec and 3.5\arcsec. When plotting this disk albedo versus the wavelength of the filter (Figure \ref{fig:disk_color}), we obtain a diagnostic tool that describes the disks geometric albedo and colors. Throughout this paper, we will apply this same procedure to the synthetic model images as well.

As can be seen from figure \ref{fig:disk_color}, the disk is relatively faint compared to the star and quite red over the entire wavelength range. Even though HD 100546 is one of the brightest disk to be observed in terms of absolute flux, it reflects only about a percent of the incoming light in the near infrared, and about a factor of 4 less in the bluest filter. Any dust model invoked to explain scattered light images must be able to explain both the relative faintness and red colors. However, as mentioned before, the geometric albedo is a product of both dust properties \textit{and} disk geometry. Therefore a geometric model of the disk is necessary to isolate the effect of the dust albedo.

\subsection{Disk model}\label{sec:diskmodel}
To interpret the scattered light images of \object{HD 100546}, we will use the 2D radiative transfer code\footnote{Since both scattering and absorption take place in 3D, this model can also be referred to as an axisymmetric 3D radative transfer code.} MCMax \citep{2009A&A...497..155M} with anisotropic scattering for the dust as described in section \ref{sec:method}. This code has been succesfully applied for modelling several observables of protoplanetary disks \citep[e.g.][]{2011A&A...528A..91V} and includes anisotropic scattering \citep{2012A&A...537A..75M}. 

In constructing the disk model, we do not have to start from scratch, as the disk geometry for HD 100546 is already well constrained by previous observations and models. We will use the SED fit presented by \cite{2011A&A...531A..93M} as a starting point for our model. The model features a depleted inner disk starting at 0.25 AU \citep{2010A&A...511A..75B,2011A&A...531A...1T} which is fairly small ($<$1 AU, \citealt{Panic:2012un}), has a large empty gap \citep{2003A&A...401..577B} and a massive outer disk starting from 13 AU to 400 AU \citep{2007ApJ...665.1391G,2010A&A...519A.110P}.

\begin{figure*}[ht]
  \includegraphics[width=0.49\textwidth]{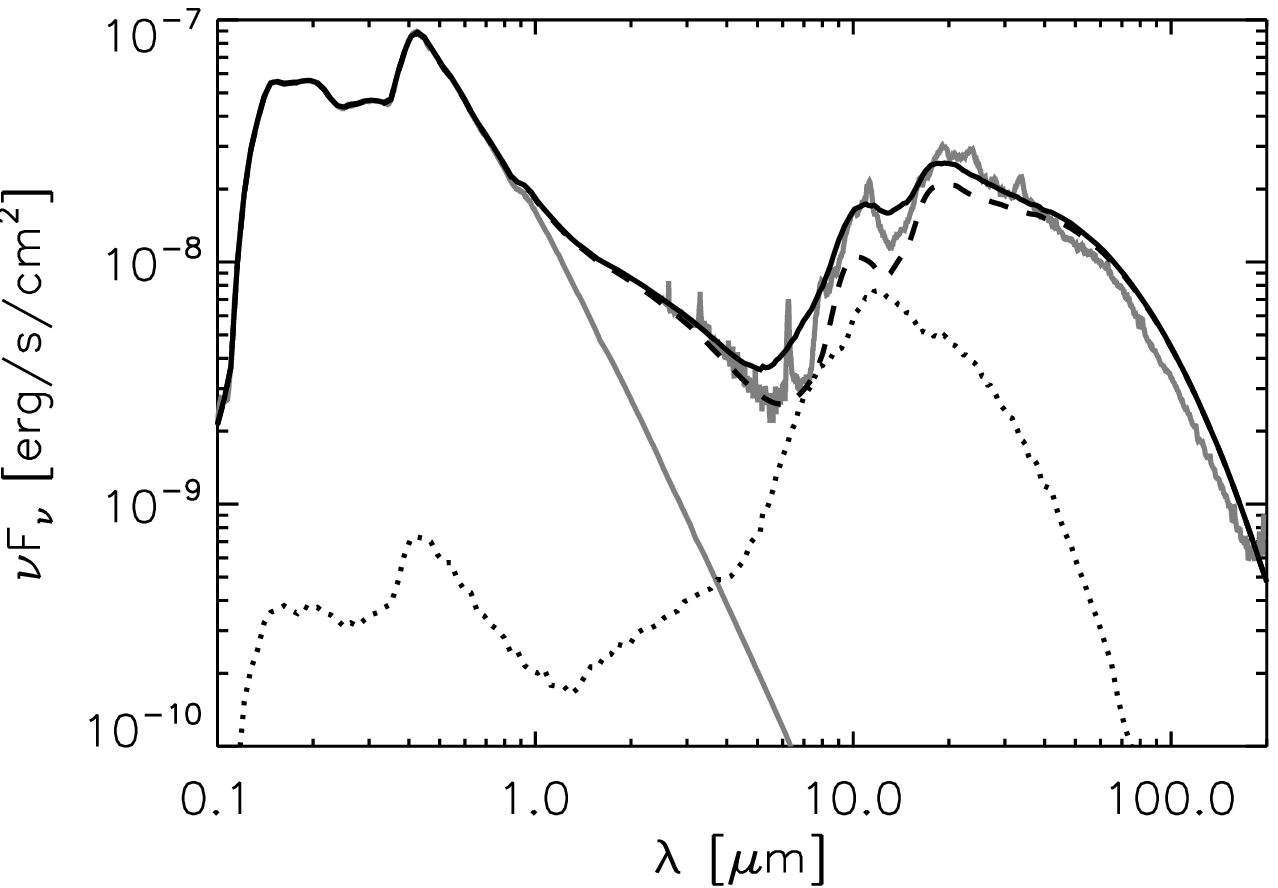}
  \includegraphics[width=0.49\textwidth]{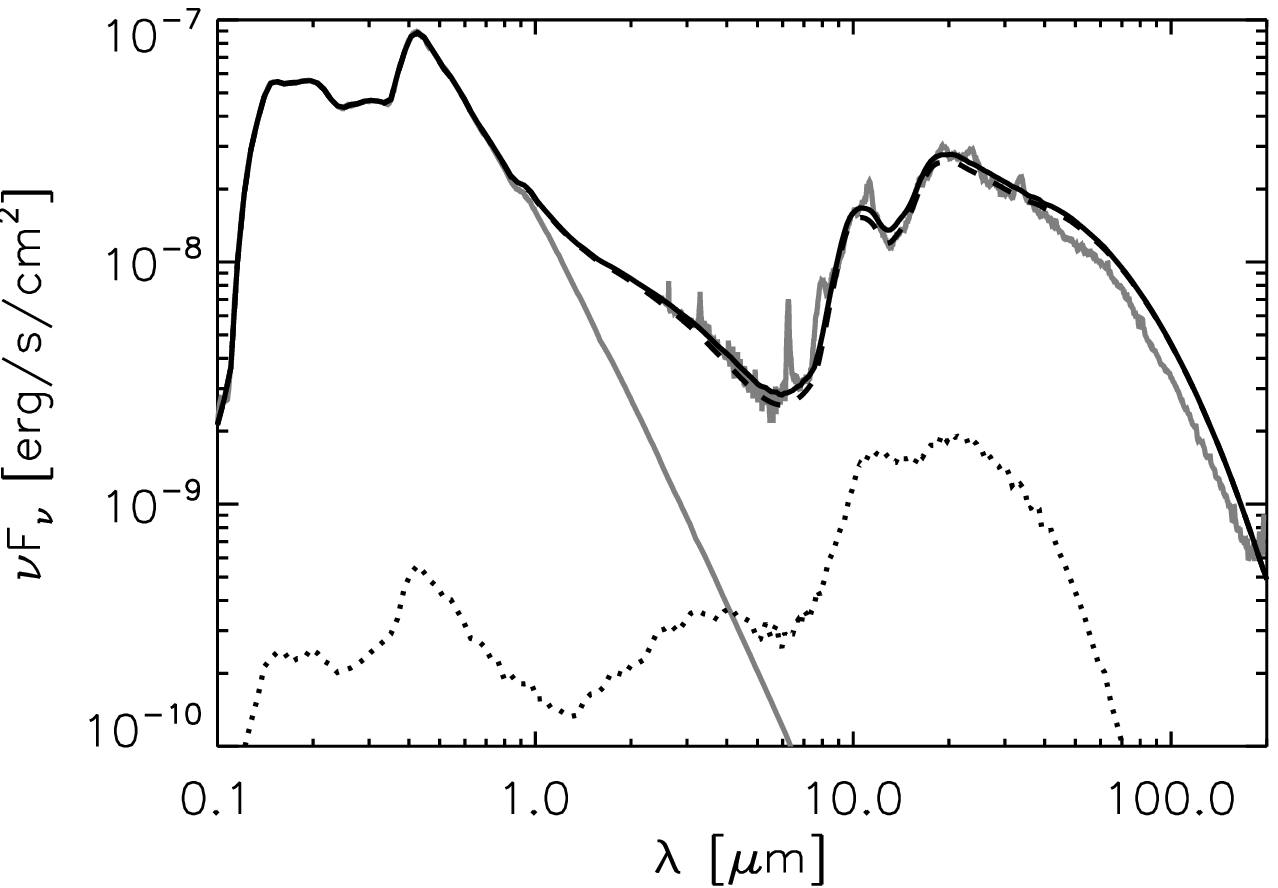}
  \caption[]{Observed SED of HD 100546 plus photosphere (grey lines). The solid lines are two model SEDs with our best-fit aggregate in the entire outer disk (left) and with the aggregates only outside of 20 AU (right). The dotted and dashed lines denote the contributions to the SED of scattered light and thermal emission, respectively .
    \label{fig:SED}}
\end{figure*}

The vertical density structure of the disk is described by a Gaussian ($\rho \propto e^{-z^2/2H_{\rm p}^2}$), where the fitted scale height is parametrized as $H_{\rm p}=0.04 {\rm AU} (r/{\rm AU})^{1.3}$, consistent with \cite{2011A&A...531A..93M}. The reason to deviate from solving the vertical hydrostatic structure is that we want to study the effect of the dust properties in a fixed geometry, whereas in a hydrostatic disk the geometry changes with dust properties through the temperature. For this purpose it is therefore easier to use the prescribed vertical structure based on the hydrostatic model, which provides an equally good fit to the SED (Fig \ref{fig:SED}).

\begin{table}
  \centering
  \begin{tabular}{ll}
    \hline \hline
    Parameter  & Value \\
    \hline
    M$_{\rm disk}$        & 0.0001 M$_\odot$ \\
    p                   & 1.0 \\
    inclination [\degr]        & 46 \\
    \hline
    R$_{\rm in}$ [AU]    &  0.25 \\
    R$_{\rm gap,in}$ [AU]    & 0.3 \\
    R$_{\rm gap,out}$ [AU]    & 13 \\
    R$_{\rm out}$ [AU]    & 350 \\
    f$_{\rm inner}$   & 0.05 \\
    \hline
    H$_{1 \rm AU}$ & 0.04 \\
    $\beta$ & 1.3 \\
    \hline \hline
  \end{tabular}
  \caption{Disk parameters for the geometrical model. The surface density profile is defined as $\Sigma(r)\propto r^{-p}$ and scaled to the total disk mass. The scaleheight is defined as $H_{\rm p}(r)= H_{1AU} r^\beta$. The gap ranges from R$_{\rm gap,in}$ to R$_{\rm gap,out}$}
\label{tab:disk}
\end{table}

The inner disk ($\lesssim$1 AU) is not directly probed by our images, but its thermal emission is a factor of a few brighter than the star in the near-infrared. Light scattered of the outer disk surface in the longest filters is therefore dominated by scattered inner disk light, rather than scattered starlight \citep{2008ApJ...673L..63P}. We take the dust composition from \cite{2011A&A...531A..93M} for the \textit{inner} disk (table \ref{tab:dust}), and keep this composition fixed if we vary the composition of the outer disk, such that the light illuminating the outer disk remains constant. A first confrontation with the observed scattered light images shows that using the same composition for the \textit{outer} disk does not work well. It overpredicts the scattered light flux by a factor of 10-30, see figure \ref{fig:disk_color} gray dotted line.

However, the albedo of this dust is very high, close to one. We first explore the effects of lowering the albedo through the dust composition to reach a lower surface brightness. We switch to a different composition which has a particularly low albedo of $\sim$0.5. This is is also the theoretical limit in the geometrical optics regime (see section \ref{sec:method}) and in practice, a lower \textit{real} albedo can only be achived by changing particle size (See Fig \ref{fig:effective}), which we will do in the next section. For an albedo of 0.5 we use a dust composition derived from a condensation sequence, as described in \cite{2011Icar..212..416M}. This lower albedo reduces the amount of scattered light significantly, but still overpredicts the observations by an order of magnitude (Fig. \ref{fig:disk_color}, dotted line). The final parameter we have to tune to obtain the low observed surface brightness is the grain size, which we will do in the next to sections.

\begin{figure}
  \includegraphics[width=\linewidth]{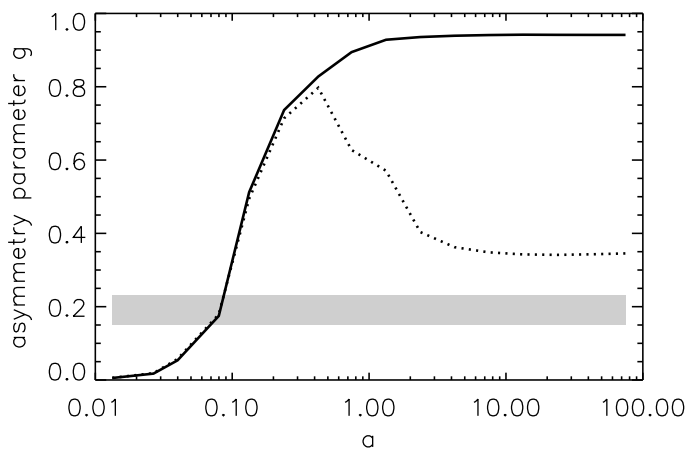}
  \caption[]{Asymmetry parameter $g$ as function of particle size at 0.6 \um, calculated in two different ways. The solid line uses the formal definition $g \equiv \langle \cos~\theta \rangle$, the dotted line is a Henyey-Greenstein function fitted to the phase function in the observed range of scattering angles. The observed range of values of the asymetry parameter is indicated by the gray area.
    \label{fig:asym}}
\end{figure}

\subsection{Brightness asymmetry}\label{sec:hg}
To constrain the characteristic grain size in HD 100546 we look first at the observed brightness asymmetry of the disk. As already explained in section \ref{sec:method}, particles smaller than the wavelength of light ($2\pi a_{\rm v} < \lambda$) scatter isotropically, whereas large particles become increasingly more anisotropic. This anisotropy can be described by the asymmetry parameter g, which is the expectation value of the cosine of the scattering angle ($g \equiv \langle cos \theta \rangle$), a parameter between 0 and 1 that increases with grain size (Fig \ref{fig:asym}).

In reality we can only sample a limited part of the phase function (34 to 126 degrees in \object{HD100546}, for a 46\degr inclination and a 10\degr{} opening angle), and $g$ can not be calculated directly. The phase function is therefore often approximated by a Henyey-Greenstein function $I_{\rm HG}(g_{\rm HG},\theta)$ \citep{1941ApJ....93...70H}, which is parametrized in such a way that the expectation value of the cosine of the scattering angle is equal to $g_{\rm HG}$. This way one can derive the asymmetry parameter with only partial knowledge of the phase function.

However, care should be taken by deriving grain sizes in this way. When grains become extremely forward scattering (g$\gtrsim 0.5$), the forward peak of the of the phase function caused by diffraction moves out of the observed range of angles (See section \ref{sec:method}, \citealt{2010A&A...509L...6M}). The observed range caused by reflection/refraction is not necessarily forward peaked (Fig. \ref{fig:phase}), and the derived value of the asymmetry parameter $g_{\rm HG}$ can become much smaller than the real asymmetry parameter $g$ (see figure \ref{fig:asym}). 

The disk of HD 100546 shows only a small brightness asymmetry of $g_{\rm HG}=0.15$ at 0.4...0.8 \um, with maxima up to $g_{\rm HG}=0.23$ \citep{2007ApJ...665..512A}. Comparing this number to the real brightness assymetry $g$ yields a grain size of $a\sim 0.08~\um$, similar to $a=0.1~\um$ derived by \cite{2000A&A...361L...9P} based on $g_{\rm HG}=0.2$. However, taking into account the limited part of the phase function observed yields a second set of solutions with grain sizes larger than one micron which come close to the observed asymmetry, see figure \ref{fig:asym}.

\subsection{Disk color and brightness}\label{sec:color}
Although small particles ($a=0.08 \um$) provide a good fit to the observed brightness asymmetries, they can not explain the disk color. They scatter in the Rayleigh limit ($2\pi a < \lambda$), and hence the scattering efficiency drops off as $\lambda^{-4}$. Such a decreasing albedo with wavelength (Fig. \ref{fig:effective}) will give rise to very blue disk colors, and it becomes clear that these small particles might explain a low albedo at one wavelength, but never the observed disk colors (Fig. \ref{fig:disk_color}, blue line). As discussed before, intermediate sized particles ($2\pi a \sim \lambda$) have a high albedo $\gtrsim 0.5$ and overpredict the disk brightness (Fig. \ref{fig:disk_color}, gray line) and also \textit{appear} too forward scattering (Fig \ref{fig:asym}).

Larger particles ($2\pi a_{\rm v} > \lambda$), on the other hand, do a much better job in explaining the disk colors and brightness. As described in section \ref{sec:method} and shown in figure \ref{fig:effective}, forward scattering decreases the \textit{effective} albedo towards shorter wavelengths, though the real albedo stays above the theoretical minimum of 0.5 (Eq \ref{eq:albedo}). A particle of 2.5 \um fits both the colors and brightness of the disk (Fig. \ref{fig:sbp} and \ref{fig:disk_color}, red lines), though its brightness asymmetry is a little too high. We will discuss this further in section \ref{sec:asym}. A particle with a different composition that has a higher \textit{real} albedo would need to be larger to reduce its \textit{effective} albedo to the same level. For example, a particle of different composition that has a \textit{real} albedo of 1.0 needs to have a size of 10 \um to explain the observed brightness.

\begin{table}
  \centering
  \begin{tabular}{lll}
    \hline \hline
    Grain & size [\um] & composition \\
    \hline
    large & 2.5 & Solar$^\dagger$ \\
    intermediate & 0.25 & Solar$^\dagger$ \\
    small & 0.08 & Solar$^\dagger$ \\
    \hline
    inner disk and wall ($<$20 AU) & 0.4 & ISM$^\dagger$ \\ 
    \hline \hline
  \end{tabular}
  \caption{Dust properties of our best fit model. 
    Opacities are calculated assuming a grain size distribution $f(a)\propto a^{-3.5}$ with typical size $a$ with a width of 0.25 dex.
    $\dagger$ Composition of
    ISM: 
    13.8\% MgFeSiO$_{4}$, 
    42.9\% MgSiO$_{3}$, 
    38.3\% Mg$_{2}$SiO$_{4}$, 
    1.8\%  NaAlSi$_{2}$O$_{6}$, 
    5\% C, 
    Solar:
    12 \% MgFeSiO$_{4}$, 
    12 \% MgFeSi$_{2}$O$_{6}$,
    12 \% Mg$_{2}$SiO$_{4}$,
    12 \% MgSi$_{2}$O$_{6}$, 
    15 \% FeS, 
    40 \% C. 
    Optical constants are from: Silicates \citep{1995A&A...300..503D,1996A&A...311..291H,1998A&A...333..188M}, Carbon \citep{1993A&A...279..577P}, Troilite \citep{1994ApJ...423L..71B}.
 }
\label{tab:dust}
\end{table}

\section{Discussion}\label{sec:Discussion}
We have shown that particles larger than 2.5 micron can explain the red colors {and low brightness in the disk of \object{HD 100546}. The question arises whether such large grains are consistent with other grain size indicators in the SED, and why we overpredict the brightness asymmetry, which we will discuss in the next sections.

\subsection{The 10 micron silicate feature}\label{sec:feat}
One grain size indicator in the SED is the 10 \um silicate feature, which indicates the presence of micron-sized grains or smaller. However, at a wavelength of 10 \um particles larger than 2.5 \um scatter extremely efficiently because they are in the resonance regime ($2\pi a \sim 10 \um$). In addition, for particles this large compared to the wavelength of 10 \um, the absorption and scattering coefficients add up to the geometrical shadow of the particle, as is the case for particles in the limit of geometrical optics (Section \ref{sec:method}). Hence the silicate feature is imprinted inversely on the scattering efficiency at this wavelength \citep{2004A&A...413L..35M}. The combination of both effects supresses the silicate feature (Fig. \ref{fig:SED}).

However, it should be noted that the silicate feature arises mainly in the disk wall \citep{2003A&A...401..577B}, whereas the scattered light images trace a region outwards of 0.5\arcsec ($\sim$50 AU at a distance of 103 pc). Radial variations in the dust properties have been observed in scattered light \citep[e.g.][Debes et al. in prep]{2011ApJ...738...23Q}, and the mineralogy and chemistry in the disk wall are also different from the outer disk \citep{2011A&A...530L...2T,2011A&A...531A..93M}. A model where the region $<$ 20 AU is dominated by small grains ($>50\%$), and the outer disk by large grains ($>99.9\%$) provides a good fit to both scattered light images and SED (Fig. \ref{fig:SED}).

It is interesting to see that this model agrees qualitatively well with the model presented in \cite{2010A&A...511A..75B} and \cite{2011A&A...531A...1T}: a disk wall dominated by small grains that produce the silicate feature, and larger grains further out. This result appears not to be in agreement with the polarimetry results from \cite{2011ApJ...738...23Q}, who find indeed a change in dust grain properties around 50...100 AU, but with small grains further out and bigger grains inwards. However, it should be noted that this conclusion is mainly based on brightness asymmetry, which we have shown might not be unambiguous in deriving grain sizes.

\begin{figure}
  \includegraphics[width=\linewidth]{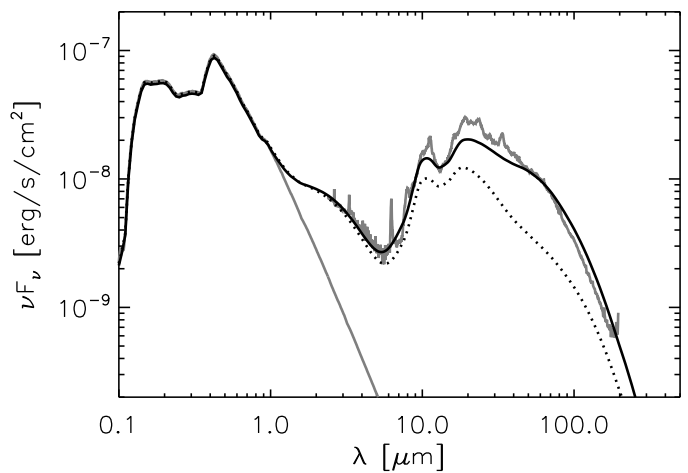}
  \caption[]{Observed SED of HD 100546 plus photosphere (grey lines). Overplotted are a series of models with selfconsistent dust settling with a turbulent mixing strength of $\alpha=0.01$, for grains with surface-to-mass-ratios of $\sigma/{\rm m}$= $10^5~\rm cm^2/g$ (solid line) and $4\cdot 10^3~\rm cm^2/g$ (dotted line), corresponding to compact grains of 0.1 and 2.5 \um, respectively.
    \label{fig:scset}}
\end{figure}

\subsection{Dust settling}\label{sec:settle}
Apart from the presence of millimeter-sized grains \citep{2003A&A...401..577B,2010A&A...511A..75B} - which are most likely located near the disk midplane - the SED of HD 100546 with its strong mid and far-infrared excess does not show strong signs of dust settling. A population of small grains in a hydrostatic disk can explain the SED \citep{2003A&A...398..607D,2011A&A...531A..93M}, showing that dust and gas must be well-mixed. In this section we explore if the large grains we find can be present at the disk surface.

To find the maximum grain size that can be present at the disk surface, we run a series of models that use self-consistent settling as described in \cite{2012A&A...539A...9M}. The dust-gas coupling in this model depends on the dust-to-gas ratio (dtg), turbulent mixing strength $\alpha_{\rm turb}$ and dust grain surface-to-mass ratio ($\sigma/{\rm m}$). For plausible values of these parameters (${\rm dtg}=0.01$, $\alpha_{\rm turb}=0.01$, we find that in a disk that has $0.005 M_\odot$ of gas a minimum surface-to-mass ratio of $10^5~\rm cm^2/g$ is required to keep particles present at the disk surface (Fig. \ref{fig:scset}). This corresponds to a 0.1 \um compact particle.

To keep particles larger than 2.5 \um present at the disk surface would require a stronger dust to gas coupling. We can think of three ways to achieve this:
\begin{enumerate}
\item A turbulent mixing strength higher than $\alpha_{\rm turb}=0.2$. Although such a high $\alpha$ could be present in the upper layers of the disk \citep[e.g.][]{2009A&A...496..597F}, it would make HD 100546 a special case among Herbig stars which have on average a much lower mixing strength \citep{2012A&A...539A...9M}, and it is beyound the scope of this paper to explore this further.
\item A gas mass higher than $0.1 M_\odot$. Although such a mass is not unreasonable for a 2.4 M$_\odot$ Herbig star \citep[e.g.][]{Williams:2011vy}, observations of gas emission lines point towards a lower rather than a higher gas mass, with a total mass in the range $0.0005...0.01 M_\odot$ \citep{2010A&A...519A.110P,2011A&A...530L...2T}.
\item Porous grains. A more porous grain has a higher surface to mass ratio than a compact particle, resulting in a stronger dust-gast coupling. Keeping the dust mass fixed, the surface-to-mass ratio of an aggregate scales with filling factor ff as
\begin{equation}
\frac{\sigma}{m}=\frac{4 \pi a^2}{m} \propto {\rm ff}^{-2/3}.
\end{equation}
because the radius scales as $a={\rm ff}^{-1/3}$. For an increase of a factor of $\gtrsim$25 in the surface-to-mass ratio this gives a filling factor of ff$\lesssim 0.01$, a very fluffy aggregate.
\end{enumerate}

Aggregates can explain the presence of large particles at the disk surface, and potentially the observed strength of the silicate feature as well \citep{2006A&A...445.1005M}. In addition, they are predicted by dust growth experiments \citep{1998Icar..132..125W,2007prpl.conf..783D} and simulations \citep[e.g.][]{2007A&A...461..215O,2008A&A...489..931Z}. 

\subsection{Phase function at intermediate scattering angles}\label{sec:asym}
However, the scattering properties of complex aggregates are less well known than those of compact spheres. The general scattering behaviour is expected to be similar for different particle structure -- i.e. a strong forward scattering peak due to diffraction and a red color of the effective albedo -- while the detailed behavior of light scattering at intermediate angles is less well established: the phase function for reflection/refraction is less well known and can be forward or backward scattering depending on the shape and structure of the particles \citep{2010A&A...509L...6M}. Since the phase function can only be derived from observations at intermediate angles, fitting the brightness asymmetry requires including the aggregate structure.

Full Discrete Dipole Approximation calculations are required to calculate agregate scattering properties, which makes them computationally impractical to use directly in radiative transfer codes. The beginnings of a computationally less expensive theory for computating aggregate opacities based on effective medium theory are there \citep{2006A&A...445.1005M}. However, effective medium theory has the disadvantage that the main assumption is that the constituents mixed are much smaller than the wavelength of incident radiation. This implies that the scattering efficiency of the constituents in this approximation is zero. We believe that in reality the constituents of the aggregates are on the order of a micron, which makes the scattering properties of the constituents an important aspect for computing the phase function of the aggregate at intermediate scattering angles. Therfore their scattering properties need to be tested before they can be applied directly to scattered light images in the way we have done in this paper.

\section{Conclusion}\label{sec:conclusion}
We have studied the effects of grain size on the colors, surface brightness and brightness asymmetry of scattered light images of circumstellar disks. We have used a 2D radiative transfer code that includes anisotropic scattering to model the disk of \object{HD 100546} over a broad wavelength range (0.4 to 2.2 \um). Our conclusions are:

\begin{itemize}
\item The low observed albedos of circumstellar dust can be explained by extreme forward scattering by grains larger than the observing wavelength. This reduces the \textit{effective} albedo below the practical lower limit of 0.5 for the \textit{real} albedo, and produces gray to red disk colors. Small grains also have low albedos, but with very blue colors.
\item The brightness asymmetry between the front and back side of a circumstellar disk is not a unique indicator of grain size. For large grains, the forward scattering peak in the phase function is not observed, while the phase function in the observed range of angles can appear more isotropic. This mimics the brightness asymmetry of small grains, but with a lower albedo.
\item The red colors, low albedo and small brightness asymmetry of the protoplanetary disk around HD 100546 can be explained by a grain size larger than 2.5...10 micron, depending on composition and particle structure. The presence of such large particles at the outer disk surface indicates they must be present in the form of porous aggregates, rather than compact particles.
\end{itemize} 

While we believe that large aggregates, rather than compact particles, are responsible for the observed emission, further research in characterizing the phase function of extremely forward scattering porous aggregates is necessary to explain their properties on basis of scattered light images.

\begin{acknowledgements}
  This research project is financially supported by a joint grant from the
  Netherlands Research School for Astronomy (NOVA) and the Netherlands
  Institute for Space Research (SRON). 
  Based on observations made with the NASA/ESA Hubble Space Telescope, obtained at the Space Telescope Science Institute (STScI), which is operated by the Association of Universities for Research in Astronomy (AURA), Inc., under NASA contract NAS 5-26555. These observations are associated with program \#'s 10167, 9295 and 9987.  Support for program \# 10167 was provided by NASA through a grant from STScI.
\end{acknowledgements}
% 
% The bibliography
% 
\bibliographystyle{aa} % style aa.bst
\bibliography{references.bib}

\end{document}